\documentstyle[11pt,newpasp,twoside]{article}
\begin{document}
\title{Hollow Core?}
\author{ G.J. Qiao$^{1,2}$, J.F. Liu$^{1,2}$, Yang Wang$^{1,2}$, 
X.J. Wu$^{1,2}$, J.L. Han$^{1,3}$}
\affil{$^1$Beijing Astrophysics Center, CAS-PKU, Beijing 100871, China\\ 
$^2$Astronomy Department, Peking University, Beijing 100871, China\\
$^3$National Astronomical Observatories, CAS, Beijing 100012, China}

\begin{abstract}
We carried out the Gaussian fitting to the profile of PSR B1237+25 and
found that six components rather than five are necessary to make a good
fit. In the central part, we found that the core emission is not filled
pencil beam but is a small hollow cone. This implies that the impact angle
could be $\beta<0.5^\circ$. The ``hollow core'' is in agreement with
Inverse Compton Scattering model of radio pulsars.
\end{abstract}

\section{Introduction}
The mean profiles are believed to be the cross-cut of the emission
windows of pulsars. Early days, the conal double profiles helped to
establish the hollow cone beam model. More components of accumulated
profiles led Backer (1976) to suggest that pulsar emission beams are
composed of a hollow cone and a central beam. Rankin(1983) further
pointed out there exist two distinct types of emission components:
one central (filled) core component and two conical components.
To seperate emission components in mean pulse profiles,
Wu et al. (1992) proposed to use Gaussian fitting to the mean
profile. This approach was developed by many authors, e.g.  
Kramer et al. (1994), Kuzmin \& Izvekova (1996). 
Here we discuss
the Gaussian components of PSR B1237+25 and conclude that the core
is not filled but is hollow.

\section{PSR B1237+25}
PSR B1237+25 was the prototype of five component pulsars (e.g. Rankin 
1983). Yet at some frequencies observations apparently show six components.
The Gaussian-fitting to the average profile gaves quantitatively the
parameters of 6 individual emission components of PSR B1237+25. In our
work, some
profiles were taken from Phillips et al. (1992) and Barter et al. (1982)
(at frequencies 130MHz, 320MHz, 430MHz, 610MHz, 1418MHz,and 2380MHz) and
some from Kuzimin et al. (1998) (at 100MHz, 200MHz, 400MHz, 600MHz, 1400MHz,
and 4700MHz). We found that the all decomposed components are in a
well-organized pattern, rather than randomly located. That implies
that the decomposed six components are physically the same for all 
the frequencies. The most interesting is the two components at the
central part, which demonstrates that the core is actually from a
hollow beam.

In fact, this is not the unique case. Wu et al. (1998) decomposed PSR
B2045+16 at some frequencies and also found that there are six components.
Krammer et al. (1994) fitted 6 components to 1.42 GHz profile of
PSR B1929+10. Kuzmin  \& Izvekova (1996) found that PSR B0329+54 are
best fitted by six components over a large frequency range.

Some theoretical models have been proposed to explain the core components,
for example, the Inverse Compton Scattering (ICS) model. The model 
(see Qiao \& Lin 1998 for details) can produce naturally the most complex
beam of core plus two cones, it also explains the frequency behaviors
of pulse profiles. When an observer's line of sight intersects the outer
cone, the inner cone and the hollow core which all can be naturally given
by the model, six components can be observed, almost exactly what we got
for PSR B1237+25. Using the ICS model, we found that an impact angle greater
than $0.4^\circ$ can not be able to produce the phase seperation of
the two core components $\Delta\phi$ at frequency 130 MHz and 320 MHz.

\section {Conclusions and Discussions}

Gaussian decompositions of PSR B1237+25 result in six individual components
from the all profiles at six frequencies, and we found the core beam is
actually hollow. This is exactly the case of small impact angle in the
Inverse Compton Scattering model, $\beta<0.5^\circ$. We noticed that
Lyne \& Manchester (1988) also got a similar impact angle through an
independent way. According to the ICS model, the emission region of
core is close to the surface of neutron stars, and the magnetic field
there should be dipole. Multipole field is not necessary to be included
to explain the hollow core emission.

\acknowledgements{
We are very grateful to Prof. Kuzmin, Drs. Zhang B., Xu R.X., Hong B.H.,
and Mr. Gao. X.Y., Pan J. and Wang H.G. for helpful discusses. This work
is partly supported by NSF of China, the Climbing Project-the National Key
Project for Fundamental Research of China, and by Doctoral Program
Foundation of Institution of Higher Education in China.}


\begin{references}
\reference{} Becker D., 1976, ApJ 209, 895
\reference{} Barter, N., Morris,D., Sieber,W.,\& Hankins,T.,1996, ApJ 258,776
\reference{} Kuzmin A.D., Izvekova V.A., 1996, Astronomy Letters 22, 394
\reference{} Kuzmin A.D., et al., 1998, A\&AS 127, 355 
\reference{} Lyne A.G. \&  Manchester R.N., 1988, MNRAS 234, 477
\reference{} Phillips A. \& Wolszczan A., 1992, ApJ 385, 273 
\reference{} Qiao G.J. \& Lin W.P., 1998, A\&A 333, 172  
\reference{} Rankin J.M. 1983, ApJ 274, 333
\reference{} Wu X.J., \& Manchester R.N., 1992a, in IAU Colloq. 128, p.362
\reference{} Wu X.J., et al. 1998, AJ 116, 1984
\end{references}
\end{document}